\documentclass[psfig]{aa}  

\begin{document}

\title{Monte Carlo techniques for time-dependent radiative transfer in 
3-D supernovae}


\author{L.B.Lucy}

\offprints{L.B.Lucy}

\institute{Astrophysics Group, Blackett Laboratory, Imperial College 
London, Prince Consort Road, London SW7 2AZ, UK}

\date{Received ; accepted }

\titlerunning{Time-dependent spectral synthesis}

\maketitle

\begin{abstract}

Monte Carlo techniques based on indivisible energy packets are described for
computing
light curves and spectra for 3-D supernovae. The radiative transfer is
time-dependent and includes all effects of O(v/c). Monte Carlo quantization
is achieved
by discretizing the initial distribution of $^{56}Ni$ into 
${\cal N}$ radioactive pellets. Each pellet decays with the emission of a
{\em single} energy packet comprising $\gamma$-ray photons 
representing {\em one} line from either the $^{56}Ni$ or the $^{56}Co$ decay
spectrum.
Subsequently, these energy packets propagate through the
homologously-expanding ejecta with
appropriate changes in the nature of their contained energy as they undergo
Compton scatterings and pure absorptions.

	The 3-D code is tested by applying it to a spherically-symmetric
SN in which the transfer of optical radiation is treated with a grey
absorption coefficient. This 1-D problem is separately solved using
Castor's co-moving frame moment equations. Satisfactory agreement is obtained.

	The Monte Carlo code is a platform onto which more advanced treatments
of the interactions of matter and radiation can be added. Some of
these have already been developed and tested in previous papers and are
summarized here.


\end{abstract}

\section{Introduction}
In earlier papers, Monte Carlo (MC) methods were used to construct spectral
synthesis codes
for SNe in their photospheric phases. Because the emphasis was on 
coarse analyses of the spectra of newly-discovered SNe,
major
simplifying approximations were made to minimize computing time and to 
ensure robustness. Thus, in the first code (Lucy 1987; Mazzali \& Lucy 1993),
the SN's atmosphere is spherically-symmetric and homologously expanding, 
the radiation field is stationary ($c = \infty$), continuum
formation is confined to a sharply-defined lower boundary  
(Schuster-Schwarzschild approximation), and line formation results from the
coherent scattering of this continuum as it propagates through the outer
layers.
Moreover, the stratifications of temperature, ionization and excitation are
derived not from first principles
but from formulae that approximate the effects of
dilution in the SN's extended atmosphere.

	In a subsequent paper (Lucy 1999b), two innovations significantly 
improved this code. First, coherent scattering was replaced
by downward branching as the mechanism of line formation. Secondly, a
computed spectrum's sampling errors were greatly reduced by using the
formal integral for the emergent intensity instead of simply binning the
escaping photon packets.

	Although these codes have proved their worth diagnostically,  
they cannot compute spectra for explosion models, a challenge that
must be faced if explosion mechanisms and progenitor scenarios are to be confronted with
observed spectra and light curves. Evidently, simplifying  assumptions
must be reconsidered in planning a more powerful and versatile code.

	First, the assumption of stationarity must be abandoned. As Arnett
(1980, 1982) long ago demonstrated for SNe both of types I and II, the
luminosity $L(t)$ at elapsed time $t$ 
does not in general closely approximate the instantaneous 
energy deposition rate by $\gamma$-rays emitted in radioactive decays.
Accordingly, the time-dependent
diffusion of radiant energy through the expanding ejecta
must be treated explicitly
if we wish to compute the emergent luminosity
density $L_{\nu}$ as a function of time. 

	Similarly, artificially creating radiant energy
by means of a continuum-emitting
lower boundary is no longer acceptable. The computational domain must
be the complete configuration, not just a 'reversing layer'; and 
continuum formation must be treated explicitly.  
Escaping radiant energy then derives ultimately from radioactive decays or
from the energy content of the ejecta at $t_{1}$, the time at which the
output from an explosion calculation is used to initiate the spectral
synthesis code.

	Removal of the constraint of spherical symmetry is also
highly desirable in view of accumulating observational evidence and
theoretical arguments implying that most and perhaps all SN explosions 
are significantly aspherical - see Wheeler (2004) for a recent review.

	Of the assumptions in the earlier codes, the only
one retained is that of homologous expansion. 
This requires that the explosion calculation, which of necessity includes
gas dynamics, be continued until all mass elements are to
a good approximation coasting ballistically. 

	With the assumptions to be dropped identified, the problem is
now defined: to solve the time-dependent, 3-D NLTE transfer problem for
UVOIR radiation in the
homologously expanding ejecta of a SN, given the distribution of 
mass and composition at an initial time $t_{1}$. Of course,
this problem is coupled to a corresponding
3-D transport problem
for the $\gamma$-rays emitted by radioactive isotopes. This coupling occurs
via the equations describing the deposition and degradation of $\gamma$-ray
energy.  

	In view of the magnitude of this problem, this paper is restricted to
describing an exploratory MC code 
in which the UVOIR radiation's interaction with matter is governed by
a grey absorption coefficient. This problem was previously treated by Pinto \&
Eastman (2000) in order to explore the sensitivity of type Ia light
curves to parameters.
Here the aim is to create a software platform onto which more realistic
physics
can be subsequently added and with which numerical techniques can be tested.

	Monte Carlo methods are a natural choice for this problem given
the experience with the earlier codes. Moreover, MC methods are
commonly
favoured for transport phenomena in geometrically-complex configurations
with no symmetries. Nevertheless, the conventional approach in which
the derivatives in the time-dependent transfer equation (RTE) are
approximated by
differences has already been partially implemented for 3-D SNe by 
H\"oflich (2003). With regard to the application of MC methods, the most
advanced work seems to be that of Kasen et al. (2004) who carry out 2-D
time-independent calculations to compute the observable characteristics
of a type Ia model with a conical hole.

\section{Monte Carlo techniques} 

In this section, after some preliminaries, recent developments in
MC technique that are 
relevant for a general spectral synthesis code are summarized. These remarks
are not restricted to the grey case considered in this paper. 

\subsection{Random numbers}

Random numbers are obtained with a double precision version of
the routine ran2 of Press et al. (1992). Such numbers are always denoted by
$z$, with each $z$ denoting an independent call to ran2.

\subsection{Energy packets}   

	As in earlier codes, the MC quanta are energy
packets. In general, these are referred to as ${\cal E}$-packets except when
specifying the nature of the contained energy (Paper I, Sect. 2). Thus
$r$-packets and $\gamma$-packets are monochromatic photon packets of UVOIR
radiation and of $\gamma$-rays, respectively. On the other hand,
$k$-packets and $i$-packets contain not radiation but thermal kinetic energy
and ionization energy, respectively. In addition, a full treatment of
radioactive decay and $\gamma$-ray
deposition will require consideration of $e^{-}$- and $e^{+}$-packets 
containing non-thermal electrons and positrons, respectively.

\subsection{Discretization}   

The SN is enclosed in a Cartesian grid containing $I^{3}$ identical
cubic cells. This grid expands with the ejecta so that each
cell contains a fixed parcel of matter. Physical variables do not vary 
spatially within cells.

	The expansion itself is discretized into $N$ time steps.
Expansion occurs in
instantaneous jumps at times $t_{n}$ separated by a constant value of
$\Delta \; log \; t$. During every time interval ($t_{n},t_{n+1}$),
the density $\rho$ 
of each cell is held fixed at the value predicted by homologous expansion 
at time $t_{n+1/2} = \sqrt{t_{n} \, t_{n+1}}$.

	Here, and throughout the paper, $t$ is the elapsed time since
explosion and is measured in the SN's centre-of-mass rest frame (rf).

	With this time-stepping, the calculation is divided into $N$
separate MC experiments,
each of which updates the radiation field throughout the
grid from time $t_{n}$ to $t_{n+1}$ for $n = 1,2, \ldots ,N$.

\subsection{Relativistic terms}   

Given the modest accuracy with which we can currently treat many of the
physical processes in SN ejecta, all terms of $O(v/c)$ in radiative transfer
could be neglected except for Doppler frequency shifts, which
are essential in line transfer. Nevertheless, looking forward to a time
when the relevant cross sections will be known to high precision, terms of
$O(v/c)$ are now included.

	The basic transfer calculations are carried out in the rf, but
transformations to and from the local co-moving frame (cmf) are convenient
in treating 
interactions with matter. The motion of an ${\cal E}$-packet is thus given by
\mbox{\boldmath $r$}$(t)$,     
its position as a function of time $t$ in the rf
Cartesian coordinate system with origin at the SN's centre of mass.  
When the grid attached to the ejecta expands instantaneously at times
$t_{n}$, the coordinates $\bf r$ and direction vectors
\mbox{\boldmath $\mu$} of the ${\cal E}$-packets remain unchanged.

\subsection{Energy conservation}   

In the context of this discretization of space and time, 
energy conservation implies
$I^{3}N$ constraints that the solution must satisfy. Thus, for each cell
and every time step,
the net emissivity of radiant energy per unit mass must be balanced by 
energy created
within the mass element after subtracting the increase $dU$ in its
internal energy and the $pdV$ work done by gas pressure. 

	With conventional transfer techniques, a solution satisfying this
huge number of constraints - perhaps $\sim 10^{8}$ - has to be achieved
iteratively - and convergence is never rapid. In contrast,
if the quanta in a MC calculation are {\em indestructible} and
{\em indivisible} ${\cal E}$-packets, local
energy conservation is automatically and rigorously obeyed, even if physical
variables have not converged to their final values (Lucy 1999a). 

	To a good approximation, energy conservation can be simplified to
the condition of thermal equilibrium 
by neglecting the gas terms $dU + pdV$ (Arnett 1980;
Pinto \& Eastman 2000a).
With this simplification, the net emissivity for UVOIR radiation in the
cmf is equal to the rate of energy deposition by $\gamma$-
rays. 

	Thermal equilibrium is rigorously obeyed by the MC
calculation because the effective cmf emissivity $j^{'}_{eff}(\nu)$
implied by using
indivisible ${\cal E}$-packets is subject to the integral constraint 
\begin{equation}
 4 \pi \int j^{'}_{eff}(\nu) \,  d \nu =
    4 \pi \int k^{'}_{\nu} J^{'}_{\nu} \, d \nu   +  {\cal H}^{'}_{\gamma}
\end{equation}
where  $k^{'}_{\nu}$ is the cmf absorption coefficient per unit volume, 
$J{'}_{\nu}$ is the cmf mean intensity, and ${\cal H}^{'}_{\gamma}$ is the
cmf $\gamma$-ray energy deposition rate per unit volume.
In contrast, when the RTE is solved conventionally, the cmf 
emissivity $j^{'}_{\nu}$
is given by the fundamental NLTE formulae relating the emission of photons
by bound-bound (b-b) and free-bound (f-b) transitions to the level
populations and the electron temperature. Eq. (1) is
then only satisfied asymptotically as the iterative procedure converges. 
	
	Note that, in contrast to $j^{'}_{\nu}$ derived from first
priniples, the MC emissivity
$j^{'}_{eff}(\nu)$ is not defined by mathematical formulae. Rather it is
defined operationally by the rules governing the
re-emission of absorbed ${\cal E}$-packets (Papers I, II).

\subsection{Statistical equilibrium}   

In addition to the constraints demanded by thermal equilibrium, 
an even larger number of constraints are required by statistical equilibrium.
During each time step and for each cell, the number
of excitations
of an atomic level must balance the number of de-excitations.
With accurate models for the atoms in the multi-species plasma,
the total number of levels might be $ \sim 10^{3}$. This is therefore the
number of constraints per cell per time step implied by statistical
equilibrium. Combining this with the earlier estimate of $I^{3}N$, we see
that $ \sim 10^{11}$ constraints must be satisfied in the course of computing
NLTE light curves and spectra of a 3-D SN.

	Indivisible ${\cal E}$-packets allow the
constraint of statistical equilibrium to be 
incorporated automatically and rigorously, 
even when excitation and de-excitation rates computed from basic formulae
do not balance. This is achieved with the macro-atom formalism (Paper I),
according to which
the rate of de-excitation of an excited state is not
determined by its level population but by the rates at which radiative
and collisional
processes from all other levels populate the level in question, either
directly or through the internal transitions of the macro-atom. 

	If the radiation field is computed from the RTE, errors in  
level populations, such as those present prior to
convergence, result in the non-physical creation or
destruction of radiant energy. This problem is a consequence of the generality
of the RTE equation with the NLTE emissivity $j^{'}_{\nu}$. 
Because of this generality, the NLTE
RTE could be used to follow the time-dependent relaxation to
statistical equilibrium. In contrast, with the macro-atom formalism,
spurious sources or sinks of radiant energy are
eliminated. This is achieved by effectively solving
a {\em less} general transfer
equation, namely one that incorporates the constraint of statistical
equilibrium and one, therefore, that cannot be used for a relaxation 
calculation.   

	A consequence of computing de-excitation rates with the macro-atom
machinery is that these rates and the corresponding emissivity
 $j^{'}_{eff}(\nu)$ are
insensitive to errors in the populations of the emitting
levels (Paper I, Sect. 6). In contrast, with the RTE,
errors in level populations translate directly into errors in $j^{'}_{\nu}$
and therefore in the derived radiation field. Evidently, the
macro-atom approach is more tolerant of departures from the NLTE solution
than is the conventional approach.

	As thus far developed, the macro-atom approach imposes the
constraint of thermal equilibrium simultaneously with that of statistical
equilibrium. Thus when a $k$-packet is created (Paper II, Sect. 4.3.1),
it is instantaneously eliminated (Paper II, Sect. 4.3.2). Thus, there is
no net energy transfer to or from the thermal pool. Moreover, this holds
even if the electron temperature does not imply thermal equilibrium.
Note also that if a $k$-packet is created by the de-activation of a
macro-atom of one atomic species, it may be eliminated by the activation of
a macro-atom of another species.

\subsection{Non-equilibrium }   

For SNe ejecta at late times , the assumptions of ionization and
thermal equilibrium break down (Fransson \& Kozma 1993). But this does not
preclude the use of indivisible ${\cal E}$-packets, since these are
fundamentally a means of tracking energy. Departures from equilibrium can be
modelled by allowing for the
finite life time of a non-radiative ${\cal E}$-packet before it converts back 
into an $r$-packet. Thus a $b-f$ process creates a $k$-packet or
an $i$-packet, but neither should be immediately
eliminated if the recombination time is not short compared to the elapsed
time $t$. Moreover, while awaiting elimination, the $k$-packet's energy
declines due to $p \, dV$ work.  

	If, during such non-equilibrium phases, statistical equilibrium 
remains a good approximation for the excited levels of each ion separately,
then this constraint can be imposed by introducing macro-{ions}
adapting the procedures of Paper I. 

	The above remarks strongly suggest that non-equilibrium phases
can be treated with closely similar techniques. Nevertheless, this must
be confirmed with test problems as in Paper I.

\subsection{$\Lambda$-iterations}   

The robustness of the emissivities derived with the macro-atom
formalism suggests that useful predictions of $L_{\nu}(t)$ may be obtained
without converging to the exact NLTE solution. Instead,
estimates of the excitation, ionization and temperature in each cell
could be derived using the characteristics of the local MC radiation field
as in Abbott \& Lucy (1985) and in the previous diagnostic codes (Sect. 1). 
The accuracy of this approach in computing ionization fractions has been
confirmed by Springmann \& Puls (1998) for an O-star wind. 

	Nevertheless, for definitive results, the NLTE solution
must be obtained. Fortunately, the use of indivisible ${\cal E}$-packets
and the macro-atom formalism facilitate this task. Because the thermal and 
statistical equilibrium constraints are directly incorporated into the
MC calculation,
convergence to the NLTE solution can be achieved with geometry-independent
$\Lambda$-iterations (Lucy 1999a; Paper II). Thus, by simply repeating the
process of bringing matter into thermal and statistical equilibrium with
the MC radiation field and then recomputing the latter, the solution
converges to the required equilibria. Moreover, convergence is rapid.
     
	With regard to thermal equilibrium, this success with
$\Lambda$-iterations
was initially demonstrated for a 1-D problem (Lucy 1999a). Recently,
the geometric independence of the technique has been demonstrated with
its successful application to demanding 3-D problems (Harries et al. 2004;
Kurosawa et al. 2004). With regard to thermal {\em and} statistical
equilibrium, experience is
thus far limited to a simple 1-D problem (Paper II).

\subsection{Monte Carlo estimators}   

A NLTE calculation requires the radiative rates of excitation,
ionization and heating. With the RTE, these are obtained by
numerical integration. But computing these quantities in
a MC simulation is not so straightforward. The obvious
approach is simply to count the relevant events - e.g,
photoionizations - in each cell in time
$\Delta t$, thereby deriving
the rate empirically. This is akin to a physics experiment in which
detectors are distributed throughout an apparatus to record 
events. But, though appealing, this
fails to make full use of the MC radiation field.
For example, even if no photoionizations occur in a cell, a non-zero
ionization rate must surely be derivable if it was traversed by
$r$-packets containing ionizing photons.

	To derive MC estimators of radiative rates, a summation procedure
based on volume elements is preferred (Lucy 1999) to the more obvious
choice of reference surfaces.
The basic building block from which the required estimators are
derived will now be stated in a more general form than in earlier papers
(Lucy 1999; Paper II).

	If, at a given position and time, $I_{\nu}(\theta,\phi)$ is the
specific intensity 
at frequency $\nu$ of a pencil of radiation propagating in direction
$(\theta, \phi)$, then $4 \pi I_{\nu}/c \times d \nu \: d \omega$ is the
instantaneous energy density of radiation in the frequency interval
$(\nu, \nu + d \nu)$ propagating within solid angle $d \omega$ about the
specified direction. This basic formula allows the MC 
radiation field to be converted into the conventional description
in terms of $I_{\nu}$.

	If, during $\Delta t$,
an $r$-packet of energy $\epsilon_{\nu}$ with $\nu \in (\nu, \nu + d \nu)$
propagates in a cell of volume V, and if for a time
$\delta t = \delta s/c$ its trajectory
is in the solid angle element $d \omega$, then  $\epsilon_{\nu}\delta s/c$   
is the packet's contribution to the cell's time-integrated radiant energy
in $d \nu \:d \omega$. 
Accordingly, the volume- and time-averaged
estimator of the specific intensity is given by  
\begin{equation}
 I_{\nu}\: d \nu \: d \omega = \frac{1}{4 \pi} \frac{\epsilon_{0}}{\Delta t}
            \frac{1}{V} \sum_{d \nu d \omega}
             \frac{\epsilon_{\nu}}{\epsilon_{0}} \: \delta s 
\end{equation}
Here $\epsilon_{0}$ is the reference value for the energy of the
${\cal E}$-packets (Sect. 3.2).

	Equation (2), when multiplied by the appropriate weight functions 
and integrated over $\nu$ and $\omega$, yields MC estimators for
any required property of the radiation field. Estimators of this type
were used initially to compute integrated mean intensities and 
absorption rates in a 1-D non-grey atmosphere in LTE (Lucy 1999a). Similar 
applications to dusty circumstellar envelopes 
have been reported by  Wolf (2003) and by Niccolini et al. (2003).
Estimators for the radiative rates
needed for NLTE calculations can also be derived (Paper II).

	The class of MC estimators derived from Eq. (2) can be described as
optimum and non-parametric. Optimum because they use all the MC
information and non-parametric because no
assumption is made about the radiation field. These
estimators are appropriate when the simulation is large enough that
in each $\Delta t$ every cell is traversed by many ${\cal E}$-packets.
If not, a functional form can be assumed for $I_{\nu}$ or $J_{\nu}$    - e.g.,
a dilute black body - and its parameters estimated from the limited number
of trajectories in $V$. This functional form can then be multiplied by the
appropriate weight functions to derive an estimate of the required quantity
by numerical integration.
This second procedure dampens the effects of sampling errors but,
insofar as the functional form is inexact, yields biased
estimators - i.e., ones that do not converge to their exact values as
${\cal N} \rightarrow \infty$, where ${\cal N}$ is the number of packets in
the simulation.

	An application where the extra generality provided by Eq.(2)
is essential is in constructing an estimator for
the source function in a medium with non-isotropic scattering.

\subsection{Emergent radiation}   

The emergent rf spectrum for a MC simulation can be derived simply by
counting 
escaping $r$-packets into frequency bins, with light travel-time taken into
account as in Sect. 4.2. But even in the spherical case,
large ${\cal N}$ is then required to keep sampling errors small enough
to allow a useful comparison with obseved spectra. The solution 
is to extract the source function from the simulation and
then calculate the emergent flux from the formal integral (Lucy 1999b).
In a particular 1-D case, the errors in the resulting spectrum correspond to
those of a binned spectrum from a simulation with ${\cal N}$ increased by
a factor of $\sim 320$. For a 3-D SN, where spectra for multiple
lines-of-sight must be computed, the binning option is almost useless,
and the already large gain
factor with the formal integral will be vastly increased.

\section{Gamma ray transport}

In a previous paper (Paper II) on the NLTE transfer of UVOIR radiation in a
SN envelope,
the ${\cal E}$-packets were all created at the lower boundary, thus
implicitly representing the outward diffusion of the energy released by
radioactive decays in the deeper interior.
But with the computational domain now being the entire ejecta,
the creation and transport of $\gamma$-rays must be treated explicitly. In
this section, therefore, the concept of indivisible ${\cal E}$-packets
is extended to
cover this aspect of the general spectral synthesis problem for SNe. 

	In a very recent paper, Milne et al. (2004) have compared results 
with various $\gamma$-ray transport codes and reviewed the physical processes
that must be treated. Their emphasis is on the prediction of $\gamma$-ray 
spectra rather on the powering of optical emission.

\subsection{Gamma ray lines}   

The radioactive decays $^{56}Ni \rightarrow \, ^{56}Co$ and   
$^{56}Co \rightarrow \, ^{56} Fe$ each occur with the emission of
a spectrum of
$\gamma$-ray lines - see Table 1 in Ambwani \& Sutherland (1988). Gamma-rays
of energy $E_{l}$ are emitted with probability ${\it f}_{l}$ 
when a parent nucleus decays. The total energy emitted per decay is
therefore $\sum E_{l} {\it f}_{l}$, giving $E_{Ni} = 1.728$ and 
$E_{Co} = 3.566$MeV for the  $^{56}Ni$ and  $^{56}Co$ nuclei, respectively.  
The $e$-folding times for these decays are $t_{Ni} = 8.80$ and 
$t_{Co} = 113.7$ days.

	The total $\gamma$-ray energy emitted by the decay sequence 
$^{56}Ni \rightarrow \, ^{56}Co \rightarrow \, ^{56} Fe$
in the limit $t \rightarrow \infty$ is
\begin{equation}
 E_{tot} = (E_{Ni}+E_{Co}){\cal M}_{rad}/m_{Ni}
\end{equation}
where ${\cal M}_{rad}$ is the initial mass of $^{56}Ni$, and $m_{Ni}$
is the mass of the  $^{56}Ni$ nucleus.

\subsection{Radioactive pellets}   

The initial mass of radioactive matter ${\cal M}_{rad}$ is quantized into
${\cal N}$
pellets, where ${\cal N} = E_{tot}/\epsilon_{0}$. These pellets have the
following property: they emit a {\em single}
$\gamma$-packet with cmf energy $\epsilon_{0}$ and containing $\gamma$-rays
whose cmf photon energy $E_{\gamma}^{'}$ corresponds to
{\em one} of the lines emitted in the decay sequence 
$^{56}Ni \rightarrow \, ^{56}Co \rightarrow \, ^{56} Fe$. Following this
event, a pellet is permanently inert. 

	As ${\cal N} \rightarrow \infty$, this model must yield
the correct time-dependent $\gamma$-ray line emissivities.
To achieve this, we first identify two kinds of pellet: a fraction
$E_{Ni}/ (E_{Ni}+E_{Co})$ are $Ni$ pellets that collectively emit the
$^{56}Ni$
spectrum, while the remainder are $Co$ pellets, and they account for the   
$^{56}Co$ spectrum.

	If a pellet is designated as a $Ni$ pellet, its decay time is
randomly chosen as $t_{\gamma} = -t_{Ni} \ell n \, z$;
and the emitted $\gamma$-packet is assigned to line $l$
from the $^{56}Ni$ spectrum with probability
$ E_{l} {\it f}_{l}/\sum E_{l} {\it f}_{l}$. On the other hand, a $Co$
pellet's decay time is
$t_{\gamma} =  -t_{Ni} \ell n \, z_{1} - t_{Co} \ell n \, z_{2}$,
where
$z_{1}$ and $z_{2}$ are independent random numbers from (0,1). Two terms are
required for the $Co$ pellets since each $^{56}Co$ nucleus is created in the
decay of 
$^{56}Ni$ nucleus. Having thus selected $t_{\gamma}$ for a $Co$ pellet,
we select a line from the $^{56}Co$ spectrum as with $^{56}Ni$.

	When a pellet decays, the emitted $\gamma$-packet is assigned 
a cmf direction vector \mbox{\boldmath $\mu$}$^{'}$ in accordance with
isotropic emission. Thus  
\mbox{\boldmath $\mu$}$^{'} = (sin \theta \, cos \phi, sin \theta \, sin \phi,
cos \theta)$, with $cos \theta = 1-2z$ and $\phi = 2 \pi z$. The 
corresponding rf vector \mbox{\boldmath $\mu$} is obtained with
the exact aberration formula - e.g., Castor (1972, Eq.(5)). The 
$\gamma$-packet's initial rf energy is then 
$\epsilon_{E} = \epsilon_{0}/(1- \mbox{\boldmath $\mu . v$}/c)$, 
where the local rf velocity $\mbox{\boldmath $v = r$}/t$.

	The positions of the pellets at $t_{1}$ are obtained by sampling
the distribution of $^{56}Ni$ predicted by the explosion model. Their
positions when they decay are then given by the assumption of homologous
expansion.   

	The $\gamma$-packets emitted in the 
interval $(t_{n},t_{n+1})$ are added to the $\gamma$- and $r$-packets
still propagating in the ejecta at time $t_{n}$ to form a list of
active ${\cal E}$-packets
whose trajectories are up-dated during this time step.

\subsection{Transport of $\gamma$-packets}   

In propagating through the ejecta, $\gamma$-rays 
create cascades of non-thermal electrons in multiple Compton
scatterings off free and bound electrons as well as being absorbed in
photoionizations. This redistribution of the energy of the emitted
$\gamma$-ray into numerous channels over a substantial volume can nevertheless
be modelled with indivisible $\gamma$-packets in such a way that the correct
physics emerges as ${\cal N} \rightarrow \infty$.

\subsubsection{Events}   

As a $\gamma$-packet propagates, it undergoes events, both
numerical
and physical. The numerical events are: escaping from the grid, 
reaching the surface of a cell, or coming to the end of the current time step
at $t = t_{n+1}$.
The currently-included physical events are Compton scattering and
photoelectric absorption.

	In describing how a $\gamma$-packet's trajectory is computed, it
suffices to explain how to find the next event along the trajectory of one 
packet (cf. Paper II, Sect.5).

	Given the rf position \mbox{\boldmath $r$} and direction vector
\mbox{\boldmath $\mu$} following an event,
the next event is identified by computing the distances along
the trajectory to all possible events and then selecting the event 
reached first. 
Since calculating distances
to the numerical events is trivial, we here treat only the physical events. 

	If $E$ is the rf energy of photons in a $\gamma$-packet,
the cmf energy is  
$E^{'} =  E \, (1- \mbox{\boldmath $\mu . v$}/c)$.
Accordingly, in the cmf the
$\gamma$-packet sees Compton scattering coefficient 
$\sigma^{'}(E^{'})$ and absorption coefficient
$k^{'}(E^{'})$.
But in the rf these transform to 
$\sigma_{E} = \sigma^{'}(E^{'}) (1- \mbox{\boldmath $\mu . v$}/c)$
and 
$k_{E} = k^{'}(E^{'}) (1- \mbox{\boldmath $\mu . v$}/c)$.
Thus, when the radiative transport is carried out in the rf, the
absorption, scattering (and emission) coefficients are direction-dependent
- e.g., Castor (1972, Eqs. (2) and (3)).
  
	With these rf coefficients determined, we select the distance
$\delta s$ along the trajectory at which a physical event will occur from
the standard MC formula
\begin{equation}
  (k_{E}+\sigma_{E}) \, \rho \, \delta s = -\ell n \, z
\end{equation}
and this event happens if $\delta s$ is smaller than the distances to the
numerical
events. Morever, when a physical event happens, it is a Compton scattering if
$z < \sigma_{E}/(k_{E}+\sigma_{E})$ and a photoelectic 
absorption if not. 

	If $\delta s$ is the distance to the selected event, the updated
space-time rf coordinates
are  \mbox{\boldmath $r$}$\, + \, \delta s  $\mbox{\boldmath $\mu$} and 
$t + \delta s/c$.

\subsubsection{Actions}   

Starting with numerical events, we now describe subsequent actions.

	If the $\gamma$-packet exits the grid, it escapes
to $\infty$, and attention then
turns to the next $\gamma$-packet in the list of those active during the
current time step.     
But if the $\gamma$-packet exits only the current cell and not the grid,
then it enters the
neighbouring cell and the search for the next event proceeds as above.
Finally, if the event is the end of the time step at $t_{n+1}$, 
computation of the trajectory is suspended, and the $\gamma$-packet's current
rf data string
\mbox{\boldmath $r$},$t$,\mbox{\boldmath $\mu$},$\epsilon_{E}$,$E$
is stored to await the next time step.

	With regard to physical events, the following actions are taken:
In the event of a photoelectric absorption, the ${\cal E}$-packet's
trajectory as a $\gamma$-packet terminates. It now becomes a $k$- or
$i$-packet (cf. Sect. 2.7)
with the same cmf energy $\epsilon^{'}_{E}$ as the absorbed
$\gamma$-packet.
Then, since energy storage in the gas is neglected (Sect. 2.4), this 
$k$- or $i$-packet converts immediately to an $r$-packet. In a general code,
the frequency $\nu$ of photons in this $r$-packet are determined by
continuum emission (f-b or f-f) or line emission following collisional
excitation, and these are treatable with the macro-atom formalism
(Papers I \& II). But here, with grey transport for UVOIR radiation, the
emitted $r$-packet
can be regarded as bolometric. Its cmf energy is
$\epsilon^{'} = \epsilon^{'}_{E}$ and its
cmf direction vector \mbox{\boldmath $\mu$}$^{'}$ is selected according to
isotropic emission. The
rf direction vector \mbox{\boldmath $\mu$} then follows from the
aberration formula, and the rf energy is
$\epsilon =  \epsilon^{'}/(1- \mbox{\boldmath $\mu . v$}/c)$. These
quantities together with \mbox{\boldmath $r$} and
$t$ comprise the data string required to initiate the $r$-packet's
trajectory in the 3-D code for UVOIR radiation (Sect. 4).

	In the event of a Compton scattering, we first
find the scattering angle $\Theta$ by randomly selecting
$cos \Theta$ from a look-up table of the percentiles 
of the $cos \Theta$ probability distribution as functions of the incident
$\gamma$-ray energy. 
A $\gamma$-ray scattered through angle $\Theta$ has its incident
energy $E^{'} = \tilde{E}^{'}  m_{e}c^{2} $ reduced to $f_{C} E^{'}$,
where  $f_{C} = 1/[1+ \tilde{E}^{'} \, (1-cos \Theta)]$,
with the remaining energy transferred to a Compton electon. Thus the
energy is divided in the ratio $f_{C} : 1-f_{C}$. Accordingly, since 
${\cal E}$-packets are indivisible, the $\gamma$-packet continues as a 
$\gamma$-packet if $z < f_{C}$. If not, it becomes an $e^{-}$ packet. 

	If the packet continues as a  $\gamma$-packet, its cmf energy
remains  $\epsilon^{'}_{E}$ but it now contains $\gamma$-rays of cmf
energy $f_{C} E^{'}$. To derive the rf values of these quantities,
we must select a new direction vector. In the cmf, the incident  
\mbox{\boldmath $\mu$}$^{'}_{1}$ and emergent
\mbox{\boldmath $\mu$}$^{'}$ direction vectors are such that
\mbox{\boldmath $\mu$}$^{'}_{1}$\mbox{\boldmath $ . \mu$}$^{'}$
$= cos \Theta$. The combination of this with azimuthal angle
$\Phi = 2 \pi z$, where
azimuth is defined with respect to polar direction           
\mbox{\boldmath $\mu$}$^{'}_{1}$, determines \mbox{\boldmath $\mu$}$^{'}$.  
The aberration formula then gives \mbox{\boldmath $\mu$} and so, in the rf,
the emergent $\gamma$-packet has energy
$\epsilon^{'}_{E}/(1- \mbox{\boldmath $\mu . v$}/c)$ and contains
$\gamma$-rays with photon energies
$f_{C} E^{'}/(1- \mbox{\boldmath $\mu . v$}/c)$. The next event for
this $\gamma$-packet is now searched for as described in Sect. 3.3.1.

	On the other hand, if the $\gamma$-packet converts to an
$e^{-}$ packet,
we assume in situ degradation into
a bolometric $r$-packet with cmf energy
 $\epsilon^{'} = \epsilon^{'}_{E}$. The rf
data string needed to initiate its subsequent trajectory is now
computed as for an $r$-packet created by photoelectric absorption - see
above.

\subsection{Test}   

The above treatment  
is a straightforward extension of the indivisible ${\cal E}$-packet idea to  
$\gamma$-ray transport. Nevertheless, it should still be tested against the
traditional MC treatment in which the MC quanta are photons
(e.g. Colgate et al. 1980; Ambwani \& Sutherland 1988). Such a test
has been carried out by computing $\gamma$-ray
deposition in a spherical SN. For simplicity, effects of $O(v/c)$ are
neglected and photoelectric absorption acts only as a guillotine when
$E$ drops below 100keV.  

	In one calculation, the pellets
emit a single $\gamma$-packet as described above. In the comparison
calculation, each pellet emits multiple photon packets with photon energies
$E_{l}$ and weights
$\propto E_{l} \, f_{l}$, thus representing the appropriate $\gamma$-ray line
spectrum. Each such packet then undergoes several Compton scatterings,
at each of which the energy of the Compton electron is deposited in situ.
This continues until a photoelectric absorption deposits the remaining
energy. The two deposition profiles converge to one another as    
${\cal N} \rightarrow \infty$.

\subsection{Energy deposition rate}   

In the present code, 
an explicit determination of ${\cal H}^{'}_{\gamma}$,
the heating rate due to the deposition of $\gamma$-ray energy
is not required. The assumption of local thermal
equilibrium implies that absorbed $\gamma$-packets are immediately 
re-emitted as $r$-packets (Sect. 3.3.2) and their subsequent propagation
is through matter with a temperature-independent grey absorption coefficient.
But in a more general code, the thermal history of the ejecta must be
calculated, and this requires ${\cal H}^{'}_{\gamma}$. In any case, this
quantity is needed here for the solution with moment equations (Sect. 5). 
 
	In Sect. 3.3, two deposition processes are considered. First there is
the loss
of energy to Compton electrons, which occurs at the rate
\begin{equation}
 {\cal L}^{'}_{C} = 4 \pi \int \bar{f}(E^{'}) \, \sigma^{'}(E^{'}) \,
                  J^{'}_{E^{'}} \, d E^{'}
\end{equation}
where $\bar{f}(E^{'})$ is the expected fraction of $E^{'}$ transferred to
the Compton
electron and $J^{'}_{E^{'}}$ is the cmf mean intensity at photon energy
$E^{'}$.
Second, there is the loss in photoionizations, which occurs at the rate
\begin{equation}
 {\cal L}^{'}_{I} = 4 \pi \int k^{'}(E^{'}) \, J^{'}_{E^{'}} \, d E^{'}
\end{equation}

	Estimators for these rates can be derived by applying Eq. (2)
in the cmf. But note that the transport of packets is carried out in the rf.
If $\delta s$ is the distance between consecutive events for a
$\gamma$-packet propagating 
in the rf, then, to $O(v/c)$, in the cmf, 
$\delta s^{'} = \delta s \, (1- \mbox{\boldmath $\mu . v$}/c)$. Accordingly,
since we also have
$\epsilon^{'}_{E} = \epsilon_{E} \,
(1- \mbox{\boldmath $\mu . v$}/c)$, the reqired estimators are
\begin{equation}
  {\cal L}^{'}_{C} = \frac{\epsilon_{0}}{\Delta t} \frac{1}{V} \sum
                    \bar{f}(E^{'}) \: \sigma^{'}(E^{'}) \:
             \frac{\epsilon_{E}}{\epsilon_{0}} \:  \delta s 
                       \:  (1- 2\mbox{\boldmath $\mu . v$}/c)
\end{equation}
and
\begin{equation}
  {\cal L}^{'}_{I} = \frac{\epsilon_{0}}{\Delta t} \frac{1}{V} \sum
                         k^{'}(E^{'}) \:
             \frac{\epsilon_{E}}{\epsilon_{0}} \:  \delta s 
                       \:  (1- 2\mbox{\boldmath $\mu . v$}/c)
\end{equation}
These summations are over all trajectory segments $\delta s$
in $V$. Thus there is no
restriction to packets that undergo Compton scatterings (Eq.[7]) or
photoelectric absorptions (eq.[8]) in $V$.  
Accordingly, the resulting deposition rate 
${\cal H}^{'}_{\gamma} =  {\cal L}^{'}_{C}  +  {\cal L}^{'}_{I} $ 
is more accurate than the crude estimate obtained by summing
the cmf energies of $\gamma$-packets that terminate their trajectories in
$V$.

\subsection{Gamma-ray spectra}   

A by-product of this 3-D time-dependent treatment of $\gamma$-ray transport
is the prediction of $\gamma$-ray spectra, observations of which have
potential for
detecting asymmetries in SN explosions (H\"oflich 2002; Hungerford et al.
2003). Crude spectra can be obtained
simply by binning escaping $\gamma$-packets according to their rf energies
$E$. But for high quality results, especially for the orientation-dependent
spectra of an asymmetric SN, the formal integral should be used (Sect. 2.10).

\subsection{Future improvements}   

Additional physical processes
can be readily incorporated into this ${\cal E}$-packet treatment of
$\gamma$-ray transport.

\subsubsection{Positron transport}   

Following Ambwani \& Sutherland (1988, Table 1), positrons emitted in
$^{56}Co$ are here assumed to annihilate in situ. But at late times,
the transport of positrons should be followed until they
escape or deposit their rest and kinetic energies in the ejecta. This
requires that we allow the $Co$ pellets to emit $e^{+}$ packets instead
of the $\gamma$-packets with $E = 0.511$MeV.

\subsubsection{Transport of non-thermal electrons}   

In treating Compton scattering, we assume
that Compton electrons degrade in situ. The basis for this is
the short stopping distance of MeV electrons compared to MeV
$\gamma$-rays (Colgate et al. 1980).
Nevertheless, we may eventually wish to follow 
the motion of the $e^{-}$-packets, especially at late epochs.

\subsubsection{Pair production}   

In addition to Compton scattering and photoelectric absorption,
a $\gamma$-ray with $E > 2 m_{e} c^{2}$ can tranform to an
$e^{+}-e^{-}$ pair as it passes an atomic nucleus - see, e.g.,
Ambwani \& Sutherland (1988) - with each of the pair having 
kinetic energy (ke) $E/2 - m_{e} c^{2}$. In stopping, the
positron releases energy $E/2 + m_{e} c^{2}$,
comprising its ke plus the rest energy $2 m_{e} c^{2}$ 
radiated when the positron annihilates with an ambient
electron. In contrast, stopping the electron releases only its ke.  

	Accordingly, when treated with indivisible ${\cal E}$-packets, pair
production converts a $\gamma$-packet into an $e^{+}$-packet with
probability $1/2+q$ or into an $e^{-}$-packet with probability
$1/2-q$, where $q =  m_{e} c^{2}/E^{'}$. In either case, the cmf
energy of the packet is that of the incident $\gamma$-packet
$\epsilon^{'}_{E}$. 
If in situ deposition is not assumed, the selected packet's motion is then
followed
to escape or deposition. Note that a return to the pair production event
to follow the other member of the pair is not required (cf. Paper I,
Sect. 1). 

\subsubsection{Rare radioactivites}   

If the evolution of the ejecta is followed to late epochs
($> 1000$ days),
then rare but long-lived radioactive isotopes such as
$^{57}Co, ^{44}Ti$ and $^{22}Na$ must be included (e.g., Woosley et al.
1989). This just requires additional types of pellet.

\section{Transport of UVOIR radiation}   

In a previous paper (Paper II) the NLTE transfer of UVOIR radiation in
a SN envelope of pure H was treated using and extending the macro-atom 
formalism (Paper I). Here, given the emphasis on time dependence and 3-D,
we simplify to a grey absorption coefficient. 

\subsection{Transport of $r$-packets}   

As with $\gamma$-packets, the transport of $r$-packets is carried out in a
sequence of MC simulations for the time steps
$t_{n} \rightarrow t_{n+1}$. Each
of these simulations directly follows the corresponding one for
$\gamma$-packets (Sect. 3.3).
 
\subsubsection{Sources of $r$-packets}   

The $r$-packets that propagate in the ejecta in the time interval
($t_{n},t_{n+1}$) have several sources. First are those 
that did not escape during the previous time step. For these,
we have the rf data string  
(\mbox{\boldmath $r$},$t$,\mbox{\boldmath $\mu$},$\epsilon$), and so their
trajectories can be continued from time $t_{n}$. Second are those created
during the current time step as $\gamma$-packets are eliminated.
This occurs either
by photoelectric absorption or by the stopping of Compton electrons 
(Sect. 3.3.2). In either case, the $\gamma$ transport routine provides the
rf data string
(\mbox{\boldmath $r$},$t$,\mbox{\boldmath $\mu$},$\epsilon$) 
needed to initiate their trajectories as $r$-packets.

	A third source is specific to the first time step.
Radiant energy emitted by pellets that decay at times prior to $t_{1}$
must be accounted for.
An approximate treatment is as follows:
A $\gamma$-packet emitted at $t_{\gamma} < t_{1}$ and position
\mbox{\boldmath $r$}$_{\gamma}$
is assumed to have converted to
an $r$-packet by time $t_{1}$ but to have diffused negligibly relative to
matter (position coupling) in the time interval ($t_{\gamma},t_{1}$)
so that at $t_{1}$
its rf position is \mbox{\boldmath $r$}$_{\gamma} (t_{1}/t_{\gamma})$.
This is justified by the short mean
free paths of photons in these early, high-density phases. The rf
direction vector at $t_{1}$ is computed on the assumption of isotropic
emission in the cmf.

	The energies of these packets must also be specified.
Although position coupled, they still do work on the expanding
ejecta. An $r$-packet's energy at $t_{1}$ is therefore
$\epsilon = \epsilon_{0} \, t_{\gamma}/t_{1}$.  
  
	An additional source of $r$-packets at $t_{1}$ is the radiation
generated by shock heating during the explosion.
The explosion model's prediction of this radiation can be discretized into 
$r$-packets of energy $\epsilon_{0}$, the same quantum of energy as for
the radioactive pellets (Sect. 3.2). However, such $r$-packets are
neglected in this test code.

\subsubsection{Events}

As an $r$-packet propagates, it undergoes numerical
and physical events. The numerical events are identical to those for
$\gamma$-packets (Sect. 3.3.1). The physical events are absorptions.

	In describing how a $r$-packet's trajectory is computed, it
suffices to explain how to find the next event along the trajectory of one 
packet. Given the rf data string
(\mbox{\boldmath $r$},$t$,\mbox{\boldmath $\mu$},$\epsilon$), 
the next event is identified by computing the distances along
the trajectory to all possible events and then selecting the event 
reached first. As with $\gamma$-packets, we treat only physical events. 

	Because the transport of $r$-packets is also carried out in the rf,
the anisotropy of the rf absorption coefficient must again be allowed
for,
as it was for $\gamma$-packets in Sect.3.3.1. If the grey
absorption coefficient per unit volume in the cmf is $k^{'}$, the effective
value in the rf is  
$k = k^{'}(1- \mbox{\boldmath $\mu . v$}/c)$,
and so the distance $\delta s$ to the absorption event is given by 
$k \,  \, \delta s = -\ell n \, z$.
This event happens if this is smaller than the distance to any numerical
event. With $\delta s$ thus determined, the coordinates
(\mbox{\boldmath $r$}$,t$) are updated as for $\gamma$-packets
(Sect. 3.3.1).

\subsubsection{Actions}   
 
For numerical events, the subsequent actions
correspond to those for $\gamma$-packets (Sect. 3.3.2). 

	If the event is an absorption, we assume instantaneous isotropic
re-emission in the
cmf. Accordingly, the new rf direction vector \mbox{\boldmath $\mu$} is
calculated as it was for emitted $\gamma$-packets in Sect. 3.2. 

	The energy content of the bolometric $r$-packet must also be updated.
If the incident packet has rf energy $\epsilon_{1}$ and
direction vector \mbox{\boldmath $\mu$}$_{1}$, its cmf
energy is 
$\epsilon^{'} = \epsilon_{1} (1- \mbox{\boldmath $\mu$}_{1}
\mbox{\boldmath $. v$}/c)$. 
Since this is conserved by the absorption-emission event, the updated rf 
energy is $\epsilon = \epsilon^{'}/ (1- \mbox{\boldmath $\mu . v$}/c)$.
Calculation of the post-event data string 
(\mbox{\boldmath $r$},$t$,\mbox{\boldmath $\mu$},$\epsilon$) is now
complete, and the search for the next event can begin.

\subsection{Bolometric light curve}   

Consider an $r$-packet with post-event data string  
(\mbox{\boldmath $r$},$t$,\mbox{\boldmath $\mu$},$\epsilon$) that is an
escapee from the grid. A distant observer at rest in the rf who detects this
packet records its arrival at what he perceives to be a time 
$\tau = t-\mbox{\boldmath $\mu.r$}/c$ after the explosion. Thus the
pairs $(\epsilon, \tau)$ detected by one such observer is the
data set from which he can construct the bolometric light curve of the
3-D SN as seen from his orientation. 

	In this paper, the 3-D code is tested by applying it to a
spherically-symmetric SN. Accordingly, we use {\em all} pairs
$(\epsilon, \tau)$ to construct its orientation-averaged bolometric
light curve (Sect. 6.3).

\section{Solution with moment equations} 

In order to test the 3-D code described in Sects. 3 and 4, we apply it
in Sect. 6 to
a spherically-symmetric SN. This allows the code to be tested
against a solution of the same problem obtained
by numerical integration of the time-dependent RTE.

\subsection{Castor's equations} 

Castor (1972) has given a general treatment, accurate to $O(v/c)$,
of radiative transfer in spherically-symmetric flows. In particular, he
derives the zeroth and first frequency-integrated moment equations in the
cmf. These two equations, applied to a homologously expanding flow with
grey absorption, are the basis of the comparison calculation.

	If we again neglect the contributions of the gas to energy balance
(Sect. 2.4), the zeroth moment equation is 
\begin{equation}
 \frac{d U^{'}_{R}}{d t} + \rho \frac{\partial L^{'}}{\partial {\cal M}_{r}}
     + \frac{4 U^{'}_{R} }{t} = {\cal H}^{'}_{\gamma}
\end{equation}
where the dependent variables are $ U^{'}_{R}$, the cmf energy density
of radiation, and $L^{'}$, the cmf luminosity variable. The independent
variables are elapsed time $t$ and the mass coordinate ${\cal M}_{r}$. Note
that time derivatives in the moment equations are Lagrangian. 

	The first moment equation is
\begin{equation}
 \frac{1}{c} \frac{d L^{'}}{d t} + \alpha_{1}
 \frac{\partial P^{'}_{R}}{\partial {\cal M}_{r}} +\alpha_{2}
 (3 P^{'}_{R} - U^{'}_{R})
 = -(k^{'} + \frac{2}{ct}) L^{'}
\end{equation}
where $\alpha_{1} = 16 \pi^{2} r^{4} \rho c$,
$\alpha_{2} =  4 \pi r  c$, and the third dependent variable $P^{'}_{R}$
is the cmf radiation pressure.

\subsection{Closure approximation} 
	
To make this system determinate, an approximate formula
relating the three moments must be imposed. Here we adopt Eddington's 
approximation, 
\begin{equation}
 P^{'}_{R} = U^{'}_{R}/3
\end{equation}
which is used to eliminate $P^{'}_{R}$ from Eq.(10). This leaves two
partial differential equations (PDEs)
in the two variables $U^{'}_{R}({\cal M}_{r},t)$ and
$L^{'}_{R}({\cal M}_{r},t)$.

\subsection{Boundary conditions}   

The PDEs must be solved subject to appropriate boundary conditions.
Clearly, at the SN's centre,
\begin{equation}
  L^{'}(0,t) = 0
\end{equation}
while, at the surface, Eddington's boundary condition $F(0) = 2 J(0)$ gives 
\begin{equation}
  L^{'}({\cal M},t) = 2 \pi R_{max}^{2} c  \, U^{'}_{R} ({\cal M},t)
\end{equation}
Here $R_{max}(t) = v_{max}t$, and ${\cal M}$ is the total mass of the ejecta.

\subsection{Initial conditions}   

In addition, initial conditions are required at $t = t_{1}$.
As in the MC code (Sect. 4.1.1), we assume that
diffusion of radiation relative to matter is still negligible (Sect. 4.1.1)
at $t_{1}$, and so
\begin{equation}
 L^{'}({\cal M}_{r}, t_{1}) = 0   
\end{equation}

	An initial condition for $ U^{'}_{R}$ 
is derived as follows:
For $t < t_{1}$, negligible relative diffusion of $\gamma$-rays implies
in situ deposition, so that ${\cal H}^{'}_{\gamma} = 4 \pi j^{'}_{\gamma}$,
the integrated $\gamma$-ray emissivity in the cmf. Accordingly, from Eq.(9),
the required initial condition is given by integrating the ordinary 
differential equation (ODE)
\begin{equation}
 \frac{d U^{'}_{R}}{d t} +
      \frac{4 U^{'}_{R} }{t} = 4 \pi j^{'}_{\gamma}
\end{equation}
which can be done analytically.

	If $f({\cal M}_{r})$
is the mass fraction of $^{56}Ni$ in the mass shell ${\cal M}_{r}$ at
$t = 0$, then  
\begin{equation}
   4 \pi j^{'}_{\gamma} = S(t) \, f \, \rho \, / \, m_{Ni} 
\end{equation}
where $S(t)$, the energy released per $^{56}Ni$ nucleus, is given by 
\begin{equation}
  S(t) = \frac{E_{Ni}}{t_{Ni}} \, e^{-t/t_{Ni}} +
     \frac{E_{Co}}{t_{Co}-t_{Ni}} \, (e^{-t/t_{Co}} - e^{-t/t_{Ni}}) 
\end{equation}

	Since each term in Eq. (17) has the same form and Eq. (15) is
linear, we consider a single exponential - i.e., 
$S(t) = E_{*}/t_{*} \, e^{-t/t_{*}}$, and we note also that homologous
expansion
implies that $\rho \propto 1/t^{3}$. Integration of Eq.(15) then gives
\begin{equation}
 U^{'}_{R} = E_{*} \,
    \left( \, \frac{t_{*}}{t} - [1+  \frac{t_{*}}{t}] e^{-t/t_{*}} \right)
    \, f \, \rho \, / \, m_{Ni} 
\end{equation}
This solution can be applied to each term in Eq. (17) to construct the
complete solution. But with $t_{1} \ll t_{Ni}$, the first ($^{56}Ni$)
term suffices.

	Note that, in deriving Eq.(18), we assumed  
$U^{'}_{R}({\cal M}_{r},t) = 0$ at $t = 0$. This is consistent with the
neglect of shock heating in the MC code (Sect. 4.1.1).

\subsection{Solution technique}   

Eqs. (9) and (10) are solved in the same way as are the equations of stellar
evolution when the $\dot{P}$ and $\dot{T}$ terms are included in
the energy equation (e.g., Schwarzschild 1958, p. 100).
Thus, the time derivatives of $U^{'}_{R}$ and
$L^{'}$ are replaced by {\ backward} difference formulae. Then, since
the solution at earlier time steps is known, the PDE problem is
effectively simplified to a two-point boundary value problem for
ODEs. When the space derivatives in the ODEs
are approximated by difference formulae, the resulting algebraic system can
be solved with an
elimination procedure (Henyey method). Here, a slight generalization
of the procedure described  
by Richtmyer (1957, pp. 101-104) has been followed.

	In stellar evolution codes, it is still current practice
(A. Weiss, private communication) to approximate the time derivative of a
variable $Q$ at time $t_{n}$ with the formula
\begin{equation}
  \left(  \frac{d Q}{d t} \right)_{n} \approx \;
                     \frac{\Delta Q_{n-1}}{\Delta t_{n-1}}
\end{equation}
where $\Delta$ denotes the forward difference operator, so that
$\Delta t_{n-1} = t_{n} - t_{n-1}$. This formula is not
centred and thus has error 
$O(\Delta t)$. But higher accuracy can be achieved by using more than one
previous model - e.g, Richtmyer (1957, p.94, item 9).
If the two previous models at $t_{n-1}$ and $t_{n-2}$ are used,
the appropriate difference approximation is obtained by fitting a
parobola and then evaluating its derivative at $t_{n}$. The resulting
formula is   
\begin{equation}
  \left(  \frac{d Q}{d t} \right)_{n} \approx \;
                    (1+a) \, \frac{\Delta Q_{n-1}}{\Delta t_{n-1}}
                     -a \, \frac{\Delta Q_{n-2}}{\Delta t_{n-2}}
\end{equation}
where
$a = \Delta t_{n-1}/( \Delta t_{n-1} + \Delta t_{n-2})$. Thus, with 
constant time steps, $a = 1/2$. The gain in accuracy is investigated in Sect.
6.2.

	In starting the integration of the PDEs, the theory of Sect. 5.4
is used to provide the previous solutions required by
Eqs. (19) and (20) as well as initial estimates of $U^{'}_{R}$ and $L^{'}$
for the first Henyey iteration.

\subsection{Light curve}   
 
The solution of the PDEs yields the two functions $U^{'}_{R}({\cal M}_{r},t)$
and $L^{'}({\cal M}_{r},t)$.
Thus we directly get the bolometric
light curve $ L^{'}(t)$ seen by a
cmf observer at ${\cal M}_{r} = {\cal M}$, 
the surface of the SN. But the quantity of interest is the 
light curve seen by a distant observer at rest in the rf. This can be
calculated using ${\cal E}$-packets as follows:

	In a small interval $\Delta t$ at $t$, the surface emits $N$
$r$-packets
at random times $t_{n}$ in $\Delta t$. These all have cmf energy
$\epsilon = L^{'}(t)\Delta t/N$, and their direction cosines are
$\mu^{'}_{n} = \sqrt{z}$, corresponding to zero limb-darkening, an assumption
consistent with Eddington's approximations. Their direction cosines $\mu_{n}$
in the rf are given by the aberration formula, and their rf energies are then
$\epsilon_{n} = \epsilon/(1-\mu_{n} v_{max}/c)$. The distant observer
perceives the $n$th packet as having been emitted at elapsed time
$\tau_{n} = t_{n} - \mu_{n} R_{max}/c$, and so the $N$ pairs
($\epsilon_{n}, \tau_{n}$) represent the contribution of the interval       
$\Delta t$ to the bolometric light curve seen by this observer. Summing
over all intervals $\Delta t$ then gives the complete light curve
(cf. Sect. 4.2).

\section{Numerical results}   

After first investigating the accuracy of bolometric light curves for
spherical SN obtained 
with the cmf moment equations, this section uses such a light
curve to check the 3-D MC code described in Sects. 3 and 4.

\subsection{Model}   

The model SN is closely similar to that used by Pinto \& Eastman (2000) to
investigate 
the parameter sensitivity of type Ia light curves. The ejecta has
uniform density and its basic parameters
are: ${\cal M} = 1.39 {\cal M}_{\sun}$,   
${\cal M}(^{56}Ni) = 0.625 {\cal M}_{\sun}$, and
$v_{max} = 10^{4}$ km s$^{-1}$.
The $^{56}Ni$ is assumed to be
stongly concentrated in the central core. Thus, $f({\cal M}_{r}) = 1$ for  
${\cal M}_{r} <  0.5 {\cal M}_{\sun}$ and then drops linearly 
to zero at ${\cal M}_{r} =  0.75 {\cal M}_{\sun}$.

	The grey absorption coefficient for UVOIR radiation is
$k^{'}/\rho = 0.1$ cm$^{2}$ g$^{-1}$, and the photoelectric absorption
coefficient $k^{'}_{E}$ for $\gamma$-rays is derived from Eq. (3)
of Ambwani \&    
Sutherland (1988) with nuclear charge $Z = 14$.

\subsection{Accuracy of moment solution}   

Numerical solutions of Castor's equations have been obtained in order
to test the MC code against an RTE treatment accurate to $O(v/c)$. But
the neglect of higher order terms in $v/c$ is not the only source of error
in the RTE solutions.
Larger errors may arise due to the closure
approximation (Sect. 5.2) and the difference approximation of time
derivatives (Sect. 5.5).

	The moment equations are solved on a uniformly-spaced grid
with $400$ grid points, the innermost at
$10^{-3} R_{max}$. The energy deposition rate ${\cal H}^{'}_{\gamma}$ in
Eq. (9) is derived with a 1-D version of the MC code
described in Sect. 3 using the estimators given in Eqs. (7) and (8). 
The number of radioactive pellets is ${\cal N} = 10^{7}$.

	To investigate sensitivity to the approximation of time derivatives,
two sequences of light curves are computed,
one using Eq.(19) to
evaluate $d U^{'}_{R}/d t$ and  $d L^{'}_{R}/d t$, the other using Eq.(20).
Along each sequence, $\Delta \, log \, t$ varies, thereby determining the
rate of convergence as  
$\Delta \, log \, t \rightarrow 0$.
For each light curve, $M^{max}_{bol}$, the peak rf bolometric
magnitude, is obtained by parabolic fitting.

\begin{figure}
\vspace{8.2cm}
\includegraphics{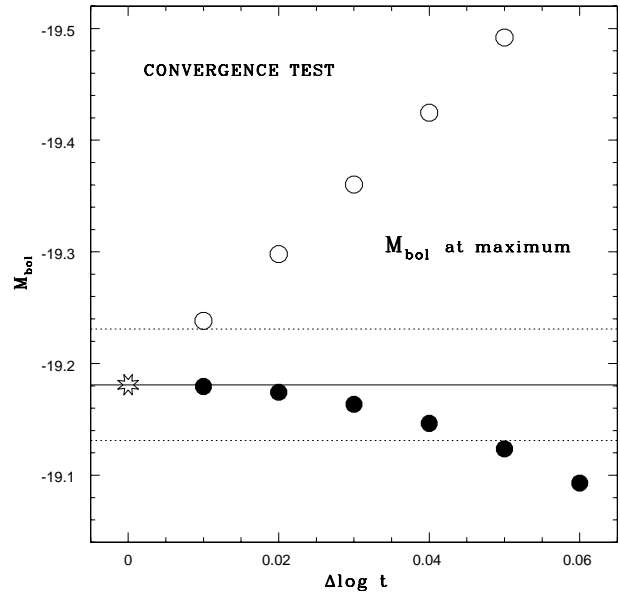}
\caption{Bolometric magnitude at maximum as function of time step
$\Delta \,log \, t$. Time derivatives are approximated with Eq. (19)
({\em open circles}) or Eq.(20) ({\em filled circles}). The extrapolation to
vanishingly small time step is shown ({\em star}) as are $\pm$ 0.05 mag. 
error bands.}
\end{figure}

	The values of $M^{max}_{bol}$ as a function of $\Delta \, log \, t$
are plotted in Fig. 1, together with the value $M^{max}_{bol} = -19.181$ 
obtained by extrapolating to infinitesimal time step. Error bands
of $ \pm 0.05$ mag. are also drawn.

	Fig. 1 shows, as expected, that errors grow linearly with Eq. (19) 
and quadratically with Eq. (20).
From this plot, we find that accuracy $< 0.05$mag. requires
$\Delta \, log \, t < 0.048$ with Eq.(20) but 
$\Delta \, log \, t < 0.009$ with Eq.(19).  Clearly,
for high precision, Eq (20) is preferred and is indeed used in Sect. 6.3.

	Although this test refers to the cmf moment equations, it is surely
relevant generally to numerical solutions of the time-dependent RTE.
Accurate difference
approximations for time derivatives are necessary to limit the
accumulation of errors as one integrates forward in time. And of course the
same remark applies to stellar
evolution codes.

\begin{figure}
\vspace{8.2cm}
\includegraphics{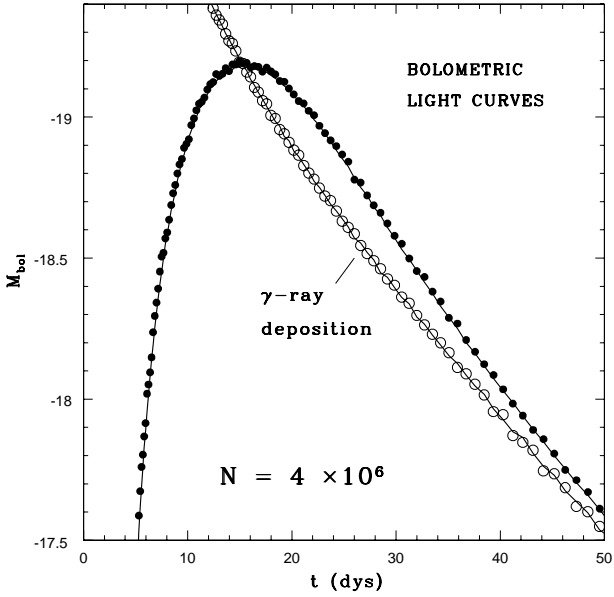}
\caption{Comparison of bolometric light curves. The light curve obtained
with the 3-D Monte Carlo code ({\em filled circles}) compared to that
obtained from the moment equations ({\em solid line}). The corresponding
$\gamma$-ray deposition curves are also plotted.}
\end{figure}

\subsection{Bolometric light curves}   

In Fig. 2, the rf bolometric light curve computed with the 3-D MC code
is compared with that derived from the cmf moment equations.
The MC light curve is from a simulation in which the initial
distribution of radioactive matter is represented by
${\cal N} = 4 \times 10^{6}$ pellets, with the resulting $\gamma$- and
$r$-packets propagating in a $100^{3}$ grid. The calculation starts at 
$log \, t_{1}(dys) = 0.3$, and the time steps are $\Delta \, log \, t = 0.01$.
The rf light curve is derived from escaping $r$-packets as described in
Sect. 4.2. Thus, for packets with arrival
times $\tau$ in the interval $(t_{n},t_{n+1})$, the values of $\epsilon$ 
are summed to obtain an estimate of $L_{n+1/2}$.  

	The rf light curve derived from the PDEs of Sect. 5.1 is also plotted
in Fig. 2. This is derived from the cmf light curve $L^{'}(t)$ as described
in Sect. 5.6. As for the calculations of Sect. 6.2, the spherical SN is
modelled with $400$ shells, and ${\cal N} = 10^{7}$ in the 1-D $\gamma$-ray
code.

	Fig. 2 shows that the two bolometric light curves are in good
agreement over an extended time interval, from well before to long after
maximum light at $t = 15.3$ days. This implies
a corresponding degree of agreement in the $\gamma$-ray energy deposition
rates. Nevertheless, these are also plotted in Fig. 2.
For the 3-D code, the deposition rate is calculated simply by summing
the energies of $\gamma$-packets as they convert to $r$-packets. But for
the 1-D code, the estimators defined by Eqs. (7) and (8) are used to
calculate ${\cal H}^{'}_{\gamma}$, which is then summed over shells.

\begin{figure}
\vspace{8.2cm}
\includegraphics{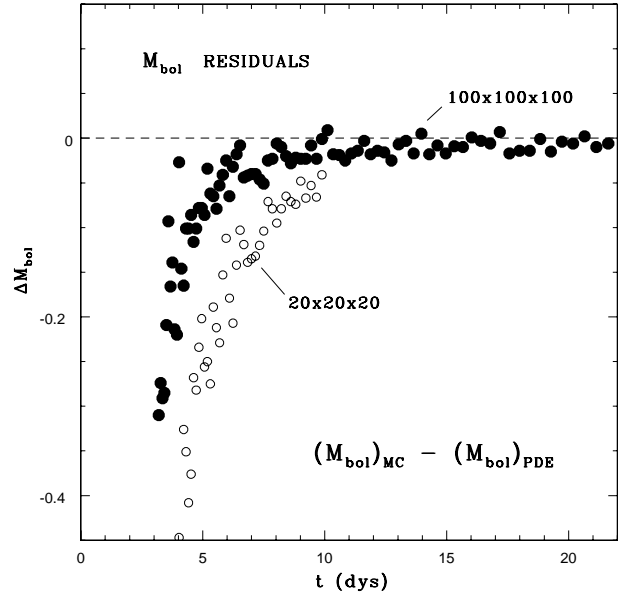}
\caption{Light curve residuals. The differences between the bolometric light
curves plotted in Fig. 2 as a function of elapsed time $t$. Residuals for
3-D MC calculations with $100^{3}$ ({\em filled circles}) and $20^{3}$ 
({\em open circles}) grids are shown.}
\end{figure}

	To investigate the precision of the 3-D light curve in more detail,
the residuals relative to the PDE solution are plotted in Fig.3.
This plot reveals large systematic residuals at early times on the rising
branch of the light curve but that these decrease with time becoming quite
small for $t \ga 8$dys. For the interval $10 - 50$ days, the mean residual  
is -0.008 mag. A residual of this order may reflect errors in the
PDE solution since this relies on Eddington's approximations.

	The large residuals on the rising branch have been traced to 
inadequate spatial resolution of the distribution of $^{56}Ni$ in the 3-D
code, despite the $100^{3}$ grid. Because $^{56}Ni$ is uniformly
distributed within each cubical cell, the function $f({\cal M}_{r})$
(Sect. 6.1) is slightly broadened relative to its more accurate
representation in the
1-D calculation. This results in some released radioactive energy reaching
the surface slightly early, giving a brighter MC light curve. This explanation
is confirmed by finding that these residuals become larger with a
coarser grid. To illustrate this, Fig. 3 includes the residuals for
$t < 10$ days with a $20^{3}$ grid.   

	This comparison with the PDE solution shows that, with a fine enough
grid, the MC
code described in Sects. 3 and 4 is capable of carrying out high precision 
transport calculations accurate to $O(v/c)$ for $\gamma$-rays and UVOIR
radiation propagating in a 3-D SN.

\subsection{Diagnostics}   

Among the merits of using ${\cal E}$-packets is conservation of energy to high 
precision and the ready monitoring of energy transformations within the
configuration.

\subsubsection{Energy conservation}   

Energy conservation for the ejecta implies that
\begin{equation}
 E_{\infty}(t) +  E_{R}(t) - W(t) =  E_{\gamma}(t)
\end{equation}
Here $E_{\gamma}(t)$ is the total energy released by radioactive decays in
the time interval $(0,t)$,  $E_{\infty}(t)$ is the total energy lost through
the surface 
in (0,t), and  $E_{R}(t)$ is the radiant energy stored in the ejecta at
time $t$, all these quantities being evaluated in the rf. Finally,
$W(t)$ is the total work done in $(0,t)$ by radiation interacting with the
expanding ejecta.

	Estimators for the quantities in Eq. (21) are sums over packets. Thus,
$E_{\gamma}(t)$ is the sum of the initial rf energies $\epsilon_{E}$ of
$\gamma$-packets emitted by radioactive pellets (Sect. 3.2);
$E_{\infty}(t)$ is the sum of the rf energies $\epsilon_{E}$ and
$\epsilon$ of $\gamma$- and $r$-packets that escape the grid in (0,t); and
$E_{R}(t)$ is the sum of the rf energies of packets still propagating within
the ejecta at time $t$. Finally, $W(t)$ is the sum 
over all physical events in $(0,t)$ of 
$\Delta \epsilon = \epsilon_{1}-\epsilon_{2}$, where     
$\epsilon_{1}$ and $\epsilon_{2}$ are an ${\cal E}$-packet's incident and
emergent rf energies, respectively.

	At all time steps in the calculation of Sect. 6.3,
the left- and right-hand sides of Eq. (21) agree to better than $1$ part in
$10^{12}$.

\begin{figure}
\vspace{8.2cm}
\includegraphics{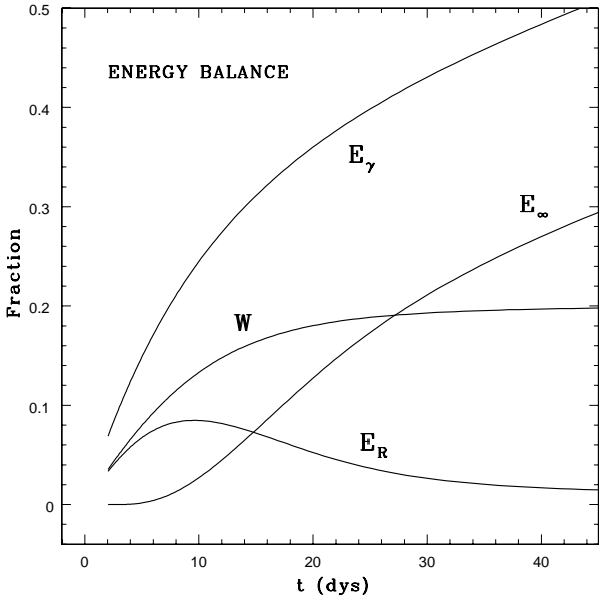}
\caption{Energy fractions as functions of elapsed time $t$. Quantities 
 plotted are $E_{\gamma}$, the integrated energy released by radioactive 
 decays, $E_{\infty}$, the integrated energy emitted to $\infty$, 
 $W$, the work done on
 the expanding ejecta, and $E_{R}$, the radiant energy stored in the ejecta.
 The unit of energy is $E_{tot}$ - see Eq. (3) - so that
 $E_{\gamma}(\infty) = 1$ .}
\end{figure}

\subsubsection{Energy flows}   

The time variations of the quantities in Eq. (21) are plotted in Fig.4,
where the unit of energy is $E_{tot}$ from Eq. (3). For the adopted 
parameters, $E_{tot} = 1.14 \times 10^{50} erg$. 

	Fig. 4 illustrates the strong departures from stationarity. 
In the early dense phases ($t \la 5$ days), energy released by radioactive
decays is trapped within the ejecta, resulting in increasing $E_{R}(t)$   
while  $E_{\infty}(t)$ remains close to zero. This continues until 
$E_{R}(t)$ reaches a maximum of $0.085$ at $t = 9.6$ days. Thereafter,
the dropping density allows the trapped radiation to be released, resulting
in a sharp increase in $E_{\infty}(t)$ and concomitant decrease in    
$E_{R}(t)$. For $t \ga 35$ days, energy storage is inconsequential  
($E_{R} \la 0.02$); stationarity then becomes a good approximation, as is
evident from the closely similar slopes of $E_{\gamma}(t)$ and 
$E_{\infty}(t)$.

	The maximum of $E_{R}(t)$ is a rough measure of the highest fraction
of ${\cal E}$-packets active at one time and for which, therefore, storage
space must be allocated. In fact, because many of the packets trapped
at $t = 9.6$ days have lost energy by doing work on the expanding ejecta,
the active fraction is larger. In the calculation reported here, the
peak active number is $7.2 \times 10^{5}$ at $t = 9.2$ days - i.e.,
$0.18 \, {\cal N}$.

	Non-stationarity is also illustrated by Fig. 5, which plots
residence against escape times for ${\cal E}$-packets in a small simulation
with ${\cal N} = 2000$. For $t \la 20$ days, we see that it is not uncommon
for escaping $r$-packets to have residence times close to the age of the SN.
Thereafter, the lowered density allows packets to escape readily and
typical residence times drop to $\sim$ a few days. Note also that no
$\gamma$-packets escape until $t \ga 20$ days.   

	In stationary problems, or this problem with $c = \infty$, a MC
simulation must continue until all packets have escaped. This limits the
usefulness of MC methods for optically thick media. But here a packet
initially at large optical depth escapes when that depth has dropped to
$\sim 1$ because of expansion. Fig. 5 illustrates this effect.

	For this problem, the difficulty with an unacceptably large number
of events before escape would reappear if $t_{1} \rightarrow 0$ were
necessary for accuracy.
Fortunately, the position coupling assumption (Sect. 4.1.1) is well 
justified for $t \la 1$ day.

\begin{figure}
\vspace{8.2cm}
\includegraphics{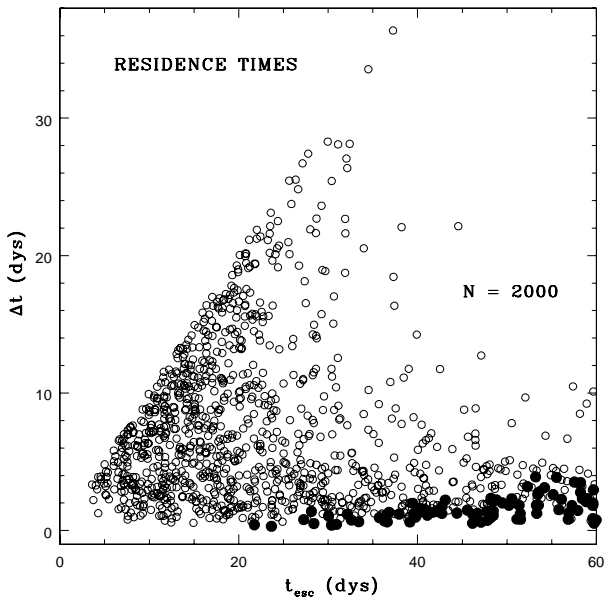}
\caption{Residence times $\Delta t = t_{esc} - t_{\gamma}$ plotted
against  $t_{esc}$. Escaping $r$ packets ({\em open circles}) and
$\gamma$-packets ({\em filled circles}) are indicated.}
\end{figure}

\section{Conclusion}   

The aim of this paper has been to initiate a MC approach to computing
light curves and spectra for 3-D SN explosions. 
The adopted procedure is based on indestructible and indivisible
${\cal E}$-packets as the MC quanta. Although
not called upon for this test problem, the attraction of this 
approach in the long-term for this and other multi-dimensional NLTE problems
is that it allows
the constraints of thermal and  
statistical equilibrium to be incorporated into the scheme via the concept
of macro-atoms. Moreover, the iterations required to achieve full
self-consistency between radiation, level populations and the electron 
temperature can be carried out with
geometry-independent $\Lambda$-iterations. 

	The calculations reported in Sect. 6 demonstrate that high
photometric precision can be achieved with MC methods for time-dependent
transport calculations in 3-D. Contributing strongly to maintaining accuracy
are the radioactive pellets introduced in Sect. 3.2 since they   
result in a seamless transition from energy transport by $\gamma$-rays to
that by UVOIR radiation. This avoids the loss of accuracy that might arise
if the $\gamma$-ray deposition profile were separately calculated and then
randomly sampled to create UVOIR radiation. An incidental benefit is
simplified coding.
 
	In Sect. 1, this investigation was described as providing a
platform onto which more detailed treatments of matter-radiation 
interactions can be added. Thus the already detailed modelling of 
$\gamma$-ray transport can be further improved as discussed in Sect. 3.7.
Also ${\cal L}^{'}_{C}$, the rate of energy deposition in the form of Compton
electrons,
is available from Eq. (7) as the source term for a Spencer-Fano
calculation of the rates at which non-thermal electrons lose energy to
thermal electrons and in ionizing and exciting atoms. Finally, the NLTE
transfer of UVOIR radiation can be implemented following the successful
test for a pure H envelope in Paper II.

	An encouraging aspect of this agenda is that the physics missing
from the present code is well understood, even though many relevant 
cross sections are still poorly known. But a discouraging aspect is
that the
demands on computer power are such that useful results for a physically-
realistic 3-D model are not feasible at present with a single workstation.
Nevertheless, as argued in Paper II, an attractive option is to implement
these
methods on a computer with numerous parallel processors. But even then,
it is probably desirable and necessary to approach the full problem
via simplified versions that are less demanding on computer time and storage
space.
Specifically, as in previous diagnostic codes, the iterations required
at each time step for full
self-consistency should initially be omitted in favour of approximate formulae
relating gas characteristics to the local MC radiation field.

\acknowledgement

I am grateful to S.A.Sim for detailed comments on the manuscript.

\end{document}